\documentclass[final]{svjour3}
\usepackage{graphicx}
\usepackage{rotating}
\usepackage{amssymb}
\usepackage{mathptmx}
\usepackage[numbers]{natbib}
\usepackage{bibentry}
\journalname{Journal of Low Temperature Physics}
\bibpunct{}{}{,}{s}{}{,}

\begin{document}

\newcommand{\hdblarrow}{H\makebox[0.9ex][l]{$\downdownarrows$}-}
\title{Benefits of bolometer Joule stepping and Joule pulsing}

\author{S. L. Stever$^{1, 2, 3}$ \and F. Couchot$^{3}$ \and V. Sauvage$^{2}$ \and N. Coron$^{2}$}
\authorrunning{S. L. Stever et al.}
% author orcid 0000-0002-3965-7080}

\institute{
\textbf{[1.]} Kavli IPMU (WPI), UTIAS, The University of Tokyo, Kashiwa, Chiba 277-8583, Japan\\
\textbf{[2.]} Institut d'Astrophysique Spatiale, \textit{INSU/CNRS}, Bt. 121, Universit\'e Paris-Sud,  Orsay, 91405, France\\
\textbf{[3.]} LAL, Univ. Paris-Sud, \textit{CNRS/IN2P3}, Universit\'e Paris-Saclay, Orsay, France\\
\email{samantha.stever@ipmu.jp}}

\maketitle

\begin{abstract}

We introduce the `Joule stepping' technique, whereupon a constantly-biased bolometer has its bias voltage modified by a small additional step. We demonstrate this technique using a composite NTD semiconductor bolometer and a pulsing device that sends an extra step in voltage. We demonstrate the results of the technique over a range of bias voltages at 100, 200, and 300 mK. Joule stepping allows us to directly measure long thermal tails with low amplitudes in the response of the global thermal architecture of bolometers, and could be a useful tool to quickly and easily calibrate the thermal time response of individual bolometric detectors or channels. We also show that the derivative of the Joule step is equivalent to the bolometer response to a $\delta$-pulse (or Joule pulse), which allows for greater understanding of transient behaviour with a better signal-to-noise ratio than pulsing alone can provide. Finally, we compare Joule step pulses with pulses produced by $\alpha$ particles, finding good agreement between their fast decay constants, but a discrepancy between their thermal decay constants.


\end{abstract}
\vspace{-2em}
\section{Introduction}
Methods for characterising bolometric detectors are important for optimisation and calibration of astronomical instruments. Bolometer responsivity is usually measured using optical techniques, but the characterisation of transient response of such detectors is more difficult. Especially for sky-scanning space telescopes, the full time response of the thermal architecture is required over a range of signal amplitudes, and the response integrals must be calibrated with a relative accuracy better than $10^{-3}$. The detection of long thermal time constants (or `tails') in astronomical instruments, especially those destined for space, can allow for calibration steps which reduce the severity of systematic effects such as cosmic rays. This issue was important for Planck HFI, where unanticipated long time constants in the $1$ second range were discovered in flight, contributing to calibration systematic errors despite their low level [1]. In the end, HFI's low-frequency instrumental response has been recovered from specific in-flight data after great effort [2]. It would have benefited from a pre-launch low-frequency characterisation, which was unavailable at the time. Therefore, access to information about long time constants in bolometric systems at the percent level is of utmost importance, especially in the era of polarisation-sensitive missions (for which miscalibration could lead to $E\rightarrow B$ leakage), and in the early stages of mission development.\\

One attempt to solve this problem is Joule stepping, in which a steadily-biased bolometer has an additional voltage step, positive and negative, injected at a chosen repetition rate. The response of the bolometer signal to this small change in the bias voltage represents the transient response of the thermal chain, and can enable longer thermal tails to be reliably measured. More importantly, the pulsing device (capable of sending stepped or pulsed signals) can be realised in a laboratory setting, and measurements can be performed quickly compared with other methods (being as precise, after one hour of data taking, as stacking of thousands of cosmic ray glitches). In principle, it could also be implemented on board as an extra in-flight calibrator.\\

In this study, we will analyse this method using an NTD semiconductor bolometer which has been well-characterised previously [3, 4, 5, 6] to verify the function of the device and attempt to characterise $\approx$1 s thermal tails, should they exist in this system. In Sec.~\ref{sec:Experiment} we will describe the pulsing scheme used. In Sec.~\ref{sec:JoulevsJoule} we will compare the response of the bolometer between Joule pulsing and Joule stepping, and in Sec.~\ref{sec:varyTandV}, we will show Joule step traces under varying temperature and bias conditions. Finally, in Sec.~\ref{sec:compa} we will compare fitted time constants of $\alpha$ particle pulses at each working point with those of the Joule step traces.\\

\vspace{-2em}
\section{Experimental setup}
\label{sec:Experiment}

\begin{figure}[htbp]
\centering
\includegraphics[width=0.7\linewidth, keepaspectratio]{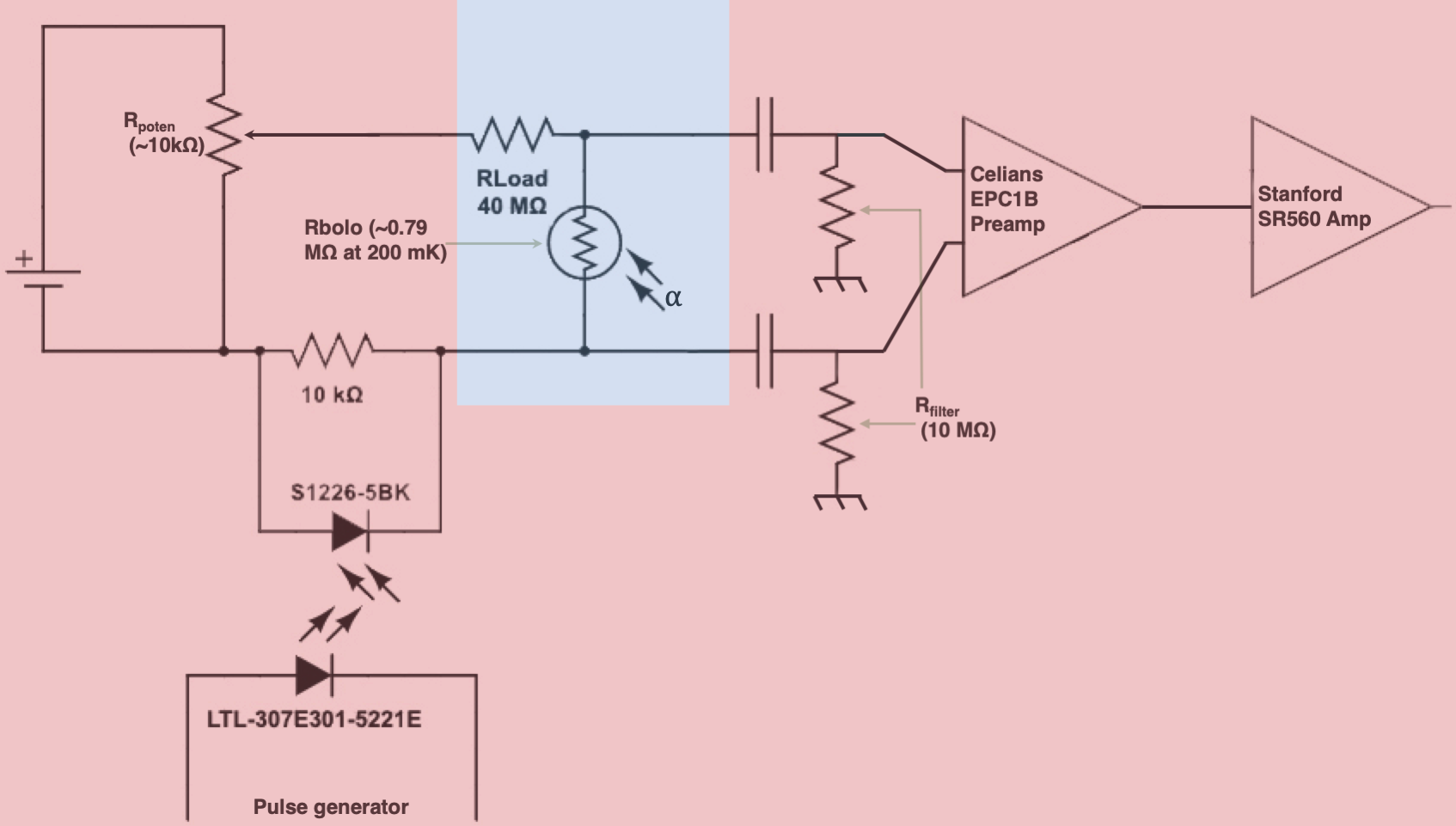}
\caption{Circuit diagram including the opto-coupled stepping device. The device takes input from a signal generator (bottom) with an output coupled to a LED. An optical fibre carries the light to the biasing box where it translates into a current in the photodiode. Blue portions represent cold components inside the cryostat, which are separated from the warm components (red) by $\approx$1.5m of cabling.  (Colour figure online)}
\label{img:schema}
\end{figure}

The NTD Ge bolometer used in this experiment is the so-called `Bolo 184' which was originally designed for use in the DIABOLO photometer [3]. The electrical circuit and experiment have been described in other papers [6], with the only changes to this setup being the cryostat in which the measurements are taken (the DRACULA cryostat at IAS [7]) and the opto-coupled stepping circuit, which we will describe here. We show the circuit diagram of the stepping device, which is integrated into the biasing circuit, in Fig.~\ref{img:schema}. A mains-powered pulse generator is used to produce square waves or short pulses with an amplitude up to $3\ $V and an adjustable repetition rate down to a fraction of a Hz. The output of the pulse generator is coupled to an Agilent blue untinted clear lens LED (model number 301-5221). The use of an optically-coupled pulsing system is an improvement upon our prior efforts to realise this device in a capacitor-coupled system in 2018 [4], in which leakage currents led to an extra bias dominating the stepping signal. By contrast, use of optical coupling allows us to prevent these ground loops and the long-term drifts from adding to the small signals measured by this experiment. An optical fibre carries the light pulse directly to the bias box, where it translates into a current in the photodiode (Hamamatsu ref.: S1226-5BK). This current flowing through the $10\ {\rm k\Omega}$ resistance gives a $\Delta V$ on top of the constant bias level. When excited by a square wave from the generator, this extra signal is well approximated by a clean step with a 10 $\mu$s rise time. This time constant is dominated by the photodiode and contains a small contribution from the capacitance of the cable feeding the bias into the cryostat. Such a small time constant is achieved owing to a small ($\approx$10$\ {\rm k\Omega}$ compared with the M$\Omega$-scale load resistor and detector) resistance acting in parallel to the high impedance bias circuit.\\

During the first tens of $\mu$s, the output shows a very fast initial rise time following the $\Delta V$ sent to the biasing circuit. One can neglect the bolometer resistance change during this phase. The new bolometer working point is then reached during a transient phase reflecting the bolometer thermal response to a variation of the Joule power in its thermistor. Since the system is linear, the signal shape during the transient phase is proportional to the integral of the pulse response shape, which allows us to directly measure the fraction of the bolometer response function corresponding to long time constants. This can be shown, for instance, from the standard pulse response (the sum of two double exponentials [8]):

\begin{equation}
f_{\textrm{glitch}}(t, t > 0) = A_1\times\left( e^{-t/\tau_{2}}-e^{-t/\tau_{1}}\right)+A_2\times\left( e^{-t/\tau_{4}}-e^{-t/\tau_{3}}\right)
\label{fiteq}
\end{equation}

\noindent where $A_{1}$ and $A_{2}$ are the amplitudes of the fast and slow components (respectively), $\tau_{1}$ is the rise time of the fast component, $\tau_{2}$ is its decay time, $\tau_{3}$ is the rise time of the slow component, and $\tau_{4}$ is its decay time (which is often considered to arise from thermal diffusion in the absorber). The integrated response reads:

\begin{eqnarray}\nonumber
f_{\textrm{step}}(t, t > 0) &  = & A_1\times\left( \tau_{2}-\tau_{1}\right)+A_2\times\left( \tau_{4}-\tau_{3}\right)\\
& \  & - \left(A_1\times\left( \tau_{2}e^{-t/\tau_{2}}-\tau_{1}e^{-t/\tau_{1}}\right)+A_2\times\left( \tau_{4}e^{-t/\tau_{4}}-\tau_{3}e^{-t/\tau_{3}}\right) \right)
\label{fiteq}
\end{eqnarray}

\noindent which allows us measure the product $A2\times (\tau_{4}-\tau_{3})/(A1\times (\tau_{2}-\tau_{1}))$ with the same signal-to-noise ratio as $A2/A1$ in the Joule pulse case. The tail extraction therefore benefits from an enhancement factor $\approx \tau_{4}/ \tau_{2}$.

\vspace{-2em}
\section{Joule Pulsing vs. Joule Stepping}
\label{sec:JoulevsJoule}

\begin{figure}[htbp]
\centering
\includegraphics[width=0.9\linewidth, keepaspectratio]{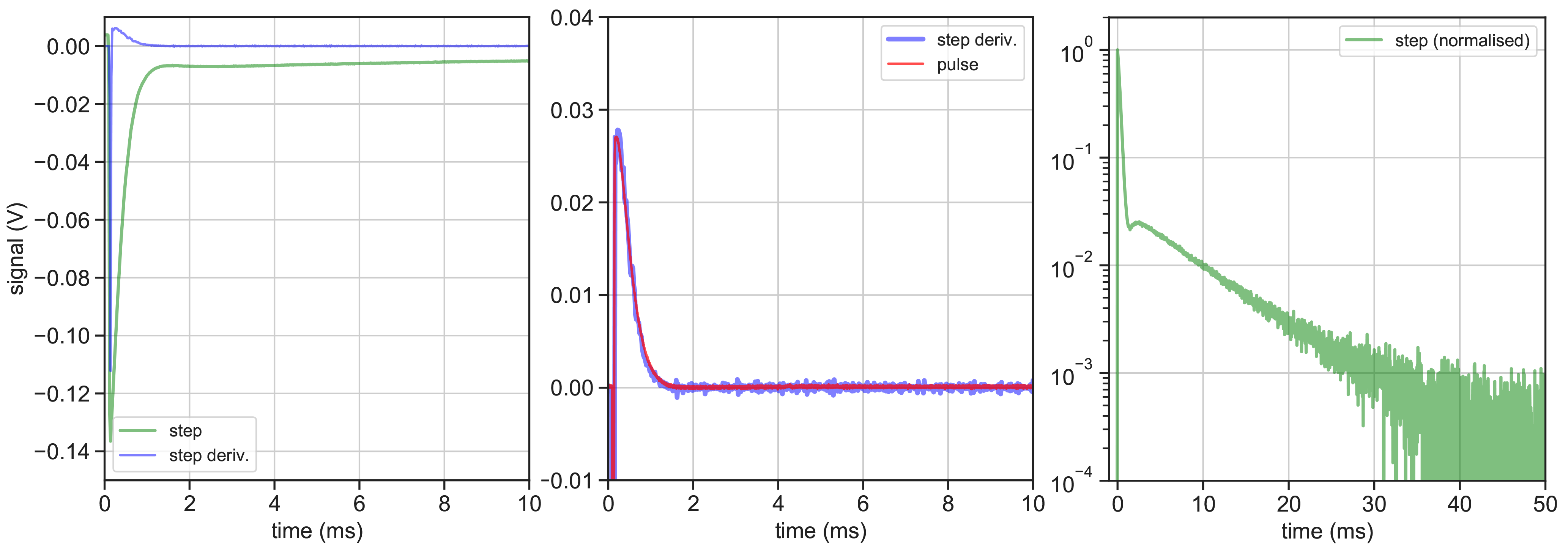}
\caption{\textit{Left}: The raw Joule step output signal (green) and its derivative (blue). \textit{Centre}: The derivative of the Joule step signal (blue) and a Joule pulse signal (red). The derivative of the Joule step signal has been rescaled to match the scale of the Joule pulse signal. \textit{Right}: The reversed normalised step response. (Colour figure online)}
\label{img:step_vs_dstep}
\end{figure}

In this section, we will assess the bolometer response to a Joule pulse (where the bias voltage has a fast $\delta_{\textrm{Dirac}}$-like step up from the constant level) and a Joule step (the small step described above). To demonstrate this, we compare one Joule step and one 40 $\mu$s Joule pulse, both at 150 mK with a bias voltage of 1 V. In Fig.~\ref{img:step_vs_dstep} (left), we show the first 10 ms of the Joule step signal, along with its derivative. We posit that the output of a Joule pulse is contained within the derivative of a Joule step, which is shown in Fig.~\ref{img:step_vs_dstep} (centre). We note that the derivative of the Joule step is normalised to match the scale of the Joule pulse, but once normalised, the output of the Joule step derivative matches that of the Joule pulse. Finally, in Fig.~\ref{img:step_vs_dstep} (right), we show the (reversed) step response normalised to the negative peak level, corresponding to the extremum of the step response curve. By normalising the Joule step traces to their peak value and considering the signal between the Joule step peak and its minimum, we can visualise the amount of information contained in the total Joule step integral. This representation allows us to directly measure the proportion of the Joule step integral which is `lost' if integration is concluded before the bolometer has fully stabilised to its new working point. In the context of time-ordered science data, this `lost' signal would contribute to the noise and measurement uncertainty, and Joule stepping allows us to measure it with a high signal-to-noise ratio compared with pulsing alone. We find that pulsing alone is not sensitive to the longer time constants, which can be seen in the stepped signals. We will demonstrate the variation of this principle in the next section.\\
\vspace{-2em}

\section{Joule stepping at varying $T$ and $V_{\textrm{bias}}$}
\label{sec:varyTandV}
Data has been taken at 100, 200, and 300 mK at various $V_{\textrm{bias}}$ between 200 and 2900 mV, which comprises a range of responsitivities and bolometer working points. At each working point, we have averaged approximately 100 Joule step traces in order to reduce noise.\\

\begin{figure}[htbp]
\centering
\includegraphics[width=0.6\linewidth, keepaspectratio]{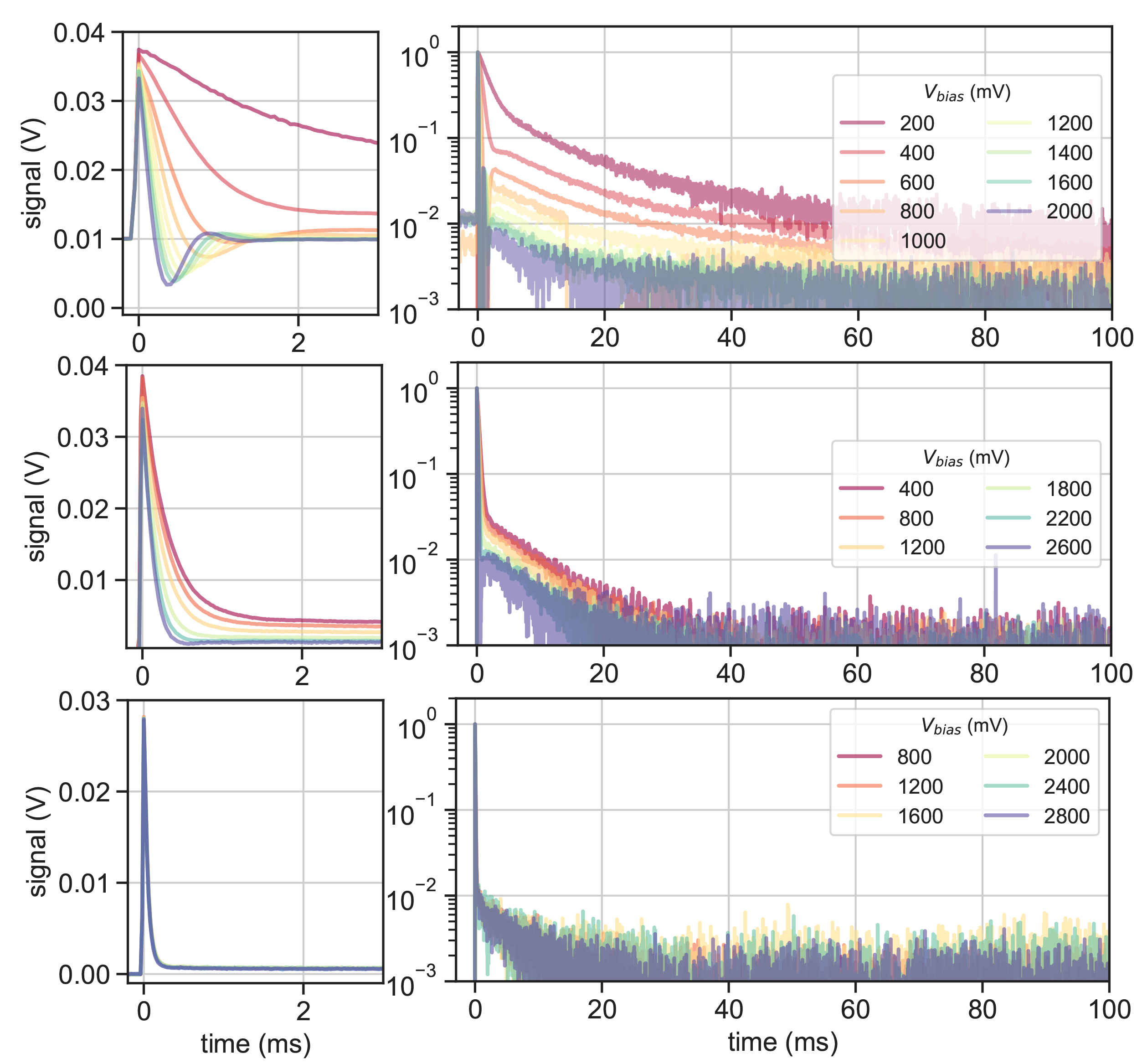}
\caption{\textit{Top}: Joule step traces at 100 mK, showing the fast component in linear scale (left) and the normalised traces in $log(y)$ scale (right). \textit{Centre}: The same at 200 mK. \textit{Bottom}: The same at 300 mK. (Colour figure online.)}
\label{img:Jstep}
\end{figure}

We show the results in Fig.~\ref{img:Jstep}, where we show the fast response of the Joule step traces in linear scale on the left side of the figure, and the slow response in normalised $log(y)$ scale. By eye, with a linear scale, there appears to be little change after the first 1-2 ms in these signals. It is only by aiding visualisation with a $log(y)$ scale that we are able to see the tails of these signals, which are $\approx$40 times longer than the short decay constants and represent $\approx$2\% of their integrals; it is therefore necessary to average at least 50-100 Joule step traces, under low noise conditions, and with a stepping period longer than any prospective thermal time constant of the system. We note some irregularities in Fig.~\ref{img:Jstep}, particularly at 100 mK, where an electro-thermal oscillation is present in the fast decay of the step trace (which we have found to be due to electro-thermal feedback and the stray capacitance of cabling at low temperatures and high $V_{\textrm{bias}}$ [4]). At 300 mK, we find little difference between the traces over various $V_{\textrm{bias}}$, which is due to the inability of the biasing circuit to provide $V_{\textrm{bias}}$ higher than 2.8 V, which is below the optimal working point of the bolometer. \\

Qualitatively, for this detector, we have demonstrated that the Joule stepping method allows us to measure the `pure' thermometer time constant, as well as to recover the global thermal time constant, which most likely arises from heating of the glue around the thermometer. The method can be used to ensure the reliability or `cleanliness' of the pulse integral, and guard against calibration biases.\\
\vspace{-3em}
\section{Comparison between Joule step traces and other fast sources}
\label{sec:compa}
In this section, we will compare the time constants obtained from the Joule step traces with those produced by another signal -- traces from $\alpha$ particles [5, 6]. To avoid biases arising from electro-thermal coupling, which can dominate the initial pulse decays in this detector at low temperatures and $V_{\textrm{bias}}$ above the ideal working point [4], we will only analyse the traces at 200 - 300 mK. To most closely approximate the effect of a Joule step in comparison with $\alpha$ particle pulses, we have taken 20 - 50 $\alpha$ glitches at each temperature and $V_{\textrm{bias}}$ combination, and averaged those with the highest proportion of their energy in the `fast' component; this is done with the intention to produce a fast injection of heat as close to the thermometer as possible. All $\alpha$ particles in this detector will be absorbed into the absorber, which has a different phonon thermal capacitance ($C_{p}$) to the NTD Ge, after which the thermal energy flows to the thermometer. By contrast, Joule steps and traces excite thermal transience directly into the NTD thermometer, with an initial rise time dominated by the photodiode in the stepping device. Therefore, we do not expect to duplicate the rise times, and expect only to find decay times within the same order of magnitude between the two methods.\\

\begin{figure}[htbp]
\centering
\includegraphics[width=0.9\linewidth, keepaspectratio]{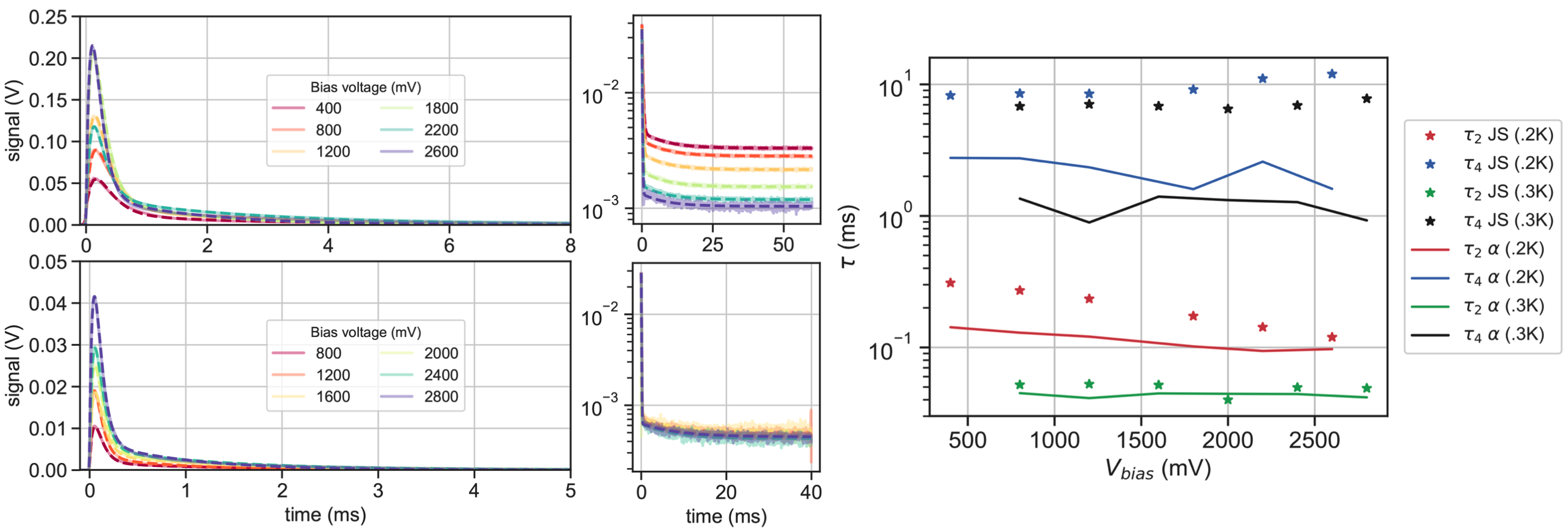}
\caption{\textit{Top row, left}: 200 mK $\alpha$ particle glitches (solid lines) and fits (dashed lines). \textit{Top row, centre}: 200 mK Joule step traces (solid lines) and fits (dashed lines). \textit{Bottom row}: $\alpha$ particle glitches and fits, and Joule step traces and fits, at 300 mK. \textit{Right}:$\tau$ as a function of $V_{\textrm{bias}}$ for the 200 and 300 mK Joule step traces (stars) and $\alpha$ particle glitches (solid lines). (Colour figure online.)}
\label{img:alphajp}
\end{figure}

Using averaged `fast' $\alpha$ glitches, we have fit over the entire glitch shape using Equation~\ref{fiteq}, as has been done in similar experiments. In the Joule step traces, we have considered the exponential fast and slow decays, which we will also call $\tau_{2}$ and $\tau_{4}$ (respectively) for ease of comparison. The fits can be seen in Fig.~\ref{img:alphajp} (left).\\

%\begin{figure}[htbp]
%\centering
%\includegraphics[width=0.6\linewidth, keepaspectratio]{taus.png}
%\caption{$\tau$ as a function of $V_{\textrm{bias}}$ for the 200 and 300 mK Joule step traces (stars) and $\alpha$ particle glitches (solid lines). (Colour figure online.)}
%\label{img:taus}
%\end{figure}

We show the time constants in Fig.~\ref{img:alphajp} (right). In the case of $\tau_{2}$, both Joule step traces and $\alpha$ glitches have decay constants in the 0.1 - 0.2 ms range at 200 mK, and 0.043 - 0.049 ms at 300 mK, showing an agreement between the two measurements. In the case of $\tau_{4}$, we note less agreement - the 200 and 300 mK $\alpha$ glitch long decay constants are shorter than those in the Joule step traces (although within the same order of magnitude). The slight disagreement between the short time constants of Joule step traces and $\alpha$ particles are due to the difference between heat slowly diffusing from the absorber (in the $\alpha$ particle case) vs. being injected directly into the thermistor (Joule step case). The discrepancy in the long time constants may relate to different heat dissipation mechanisms taking place between the two methods, although this is conjecture and and will be studied with modelling in future work.\\
\vspace{-3em}
\section{Conclusions}
We have demonstrated that the output of bolometer Joule stepping allows us to measure long time constants in bolometric detector response chains, and that the derivative of the Joule step trace is equivalent to the response function to a $\delta$ pulse in voltage. This method allows for precise measurements of very small (2\% level) contributions to the pulse integral, which has been shown to be a potential effect in sky-scanning space missions. The Joule stepping method is simple to implement, and measurements take only a few minutes, even when averaging many low-frequency ($\leq$1 Hz) signals. It also allows for in-sky calibration of detector responsivity during flight conditions, and verification of the strength of thermal coupling between various detector components.\\
\vspace{-2em}
\begin{acknowledgements}
The author wishes to acknowledge CNES for their doctoral funding during the majority of this work. This experimental work has been supported by a cofunding between CNES, Région Ile de France (DIM-ACAV) and a chaire Paris-Saclay. Kavli IPMU is supported by the World Premier International Research Center Initiative (WPI), MEXT, Japan. 
\end{acknowledgements}

\pagebreak


\begin{thebibliography}{99}

\bibitem{ade2014planck}
P.A.R. Ade, N. Aghanim, C. Armitage-Caplan et al. {\it Astron. Astrophys} \textbf{571}, A10, (2014), DOI:10.1051/0004-6361/201321577

\bibitem{adam2015data}
R. Adam, P. A. R. Ade, N. Aghanim, M. Arnaud et al. {\it Astron. Astrophys} \textbf{594}, A7, (2016), DOI:10.1051/0004-6361/201525844

\bibitem{diabolo}
A. Benoit, F. Zagury, N. Coron, M. De Petris, F-X. D\'esert, M. Giard, et al. {\it A\&AS} \textbf{141}, 3, (2000), DOI:10.1051/aas:2000129

\bibitem{samthese}
S. L. Stever {\it Ph. D. thesis}, (2019), tel-02091039

\bibitem{samalpha}
S. L. Stever, F. Couchot, N. Coron, R. M. J. Janssen, et al. {\it Proc. SPIE} \textbf{10698}, 1069863, (2018), DOI:10.1117/12.2313968

\bibitem{setupDIABOLO}
S. L. Stever, F. Couchot, N. Coron, B. Maffei  {\it J. Instrum.} \textbf{14(01)}, P01012, (2019), DOI:10.1088/1748-0221/14/01/P01012

\bibitem{reinier}
R. M. J. Janssen, S. L. Stever, V. Sauvage, G. Rouille, et al. {\it Proc. SPIE} \textbf{10709}, 107091V, (2018), DOI:10.1117/12.2311686

\bibitem{Matteo}
M. D'Andrea, A. Argan, S. Lotti, C. Macculi, et al. {\it Proc. SPIE} \textbf{9905}, 9055X, (2016), DOI:10.1117/12.2231412





\end{thebibliography}
\end{document}